\documentclass[aps,prb,twocolumn,showpacs,preprintnumbers,amsmath,amssymb]{revtex4}
\usepackage{graphicx}% Include figure files
\usepackage{dcolumn}% Align table columns on decimal point
\usepackage{bm}% bold math
\usepackage{dsfont}
\begin{document}

%\preprint{APS/123-QED}

\title{Three-Dimensionally Confined Optical Modes \\in Quantum Well Microtube Ring Resonators}% Force line breaks with \\

\author{Ch.~Strelow}
\author{C. M. Schultz}%
\author{H.~Rehberg}%

\author{H.~Welsch}%
\author{Ch.~Heyn}%
\author{D.~Heitmann}%
\author{T.~Kipp}
 \email{tkipp@physnet.uni-hamburg.de}

\affiliation{%
Institut f\"{u}r Angewandte Physik und Zentrum f\"{u}r
Mikrostrukturforschung, Universit\"{a}t Hamburg, Jungiusstra{\ss}e
11, 20355 Hamburg, Germany
}%

\date{\today}% It is always \today, today,

\begin{abstract}
We report on microtube ring resonators with quantum wells embedded
as an optically active material. Optical modes are observed over a
broad energy range. Their properties strongly depend on the exact
geometry of the microtube along its axis. In particular we observe
(i) preferential emission of light on the inside edge of the
microtube and (ii) confinement of light also in direction of the
tube axis by an axially varying geometry which is explained in an
expanded waveguide model.
\end{abstract}

\maketitle

Semiconductor microcavities confining light on the scale of its
wavelength have gained considerable interest in the last years, both
for possible applications in integrated optoelectronics and for
basic research on light-matter interaction, the latter especially in
cavity quantum electrodynamic experiments.
\cite{vahala03,khitrova06} Another growing field in the last years
is the research on semiconductor micro- and nanostructures
fabricated by use of strain relaxation of pseudomorphically grown
bilayers, which are lifted off the substrate.
\cite{prinz00,schmidt01,hosoda03,mendach05,kipp06,mendach06}

We recently have shown that, by combining both research fields, it
is possible to fabricate rolled-up microtubes which act as novel
kinds of optical ring resonators \cite{kipp06}. In this paper we
report on a new microtube ring resonator design with embedded
quantum wells (QWs) as optically active material instead of
self-assembled quantum dots as in our previous work. We observe
optical modes over a broad energy range expanding from a little
above the QW center emission energy to its low energy side. The
quality factor of the modes and its mode spacing is affected by
reabsorption of light by the QW. We study the mode structure of our
microtubes by scanning photoluminescence spectroscopy along the tube
axis. The interesting aspect of our investigation is that we clearly
observe (i) preferential emission of light at the axial edges and
(ii) confinement of optical modes also along the tube axis. The
latter is caused by spatially varying revolution numbers. Taking
into account the exact geometric dimensions which we derive from
scanning electron microscope (SEM) investigations, the spatial
dependency of the mode energies along the tube axis can be nicely
calculated.

\begin{figure}
\includegraphics{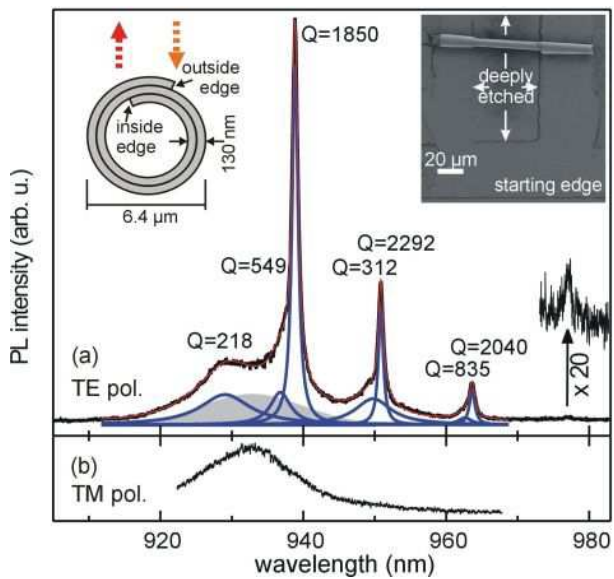}% Here is how to import EPS art
\caption{\label{fig1} (color online). Measured PL spectra (black
lines) of the microtube for (a) TE and (b) TM polarization. The
excitation power density was about 4 kW/cm$^2$. The spectrum in (a)
is approximated by the sum (red line) of Lorentzian curves for the
resonant modes (blue) and a Gaussian curve for the leaky modes (grey
shaded). Q factors have been obtained by fitting. The left inset
shows a non-scaled cross-sectional sketch of the microtube together
with the excitation and collection configuration, the right inset
shows a SEM image of the microtube. }
\end{figure}

Starting point of the fabrication of our QW microtube ring resonator
is a MBE grown layer system. On a GaAs substrate and a GaAs buffer
layer, 40 nm AlAs will serve as a sacrificial layer in the later
processing. The strained layer system which will by lifted off the
substrate consists of 14 nm In$_{0.15}$Al$_{0.21}$Ga$_{0.64}$As, 6
nm In$_{0.19}$Ga$_{0.81}$As, 41 nm Al$_{0.24}$Ga$_{0.76}$As, and 4
nm GaAs. Both In-containing layers are pseudomorphically strained
grown. The InGaAs layer forms a QW sandwiched between higher bandgap
barriers. The actual preparation process of self-supporting
microtube bridges by using optical lithography and wet etching
processes starts with the definition of a U-shaped strained mesa
followed by the definition of a starting edge. Details can be found
in Ref. \onlinecite{kipp06}. Here, we further improved our
preparation technique by etching deeply into the AlAs layer in the
region between the legs of the U-shaped mesa  [see SEM picture in
Fig. \ref{fig1}]. During the following selective etching step this
region is protected by photoresist. This process leads to a larger
and more controllable lifting of the center part of the microtube
from the substrate. The SEM image in Fig. \ref{fig1} depicts the
specific microtube on which all measurement presented in this paper
have been performed. It is slightly more than twofold rolled in its
self-supporting part, correspondingly, the wall thickness is mostly
130 nm except for the small region along the tube axis where three
strained sheets sum up to a 195 nm thick wall. The self-supporting
part has a diameter of about 6.4 $\mu$m and a distance to the
substrate of about 1 $\mu$m. The left inset in Fig. \ref{fig1}
sketches a \emph{non-scaled} cross section of the microtube.

We investigated our microtubes by micro photoluminescence
spectroscopy at low temperature $T = 7$ K. A He-Ne laser ($\lambda =
633$ nm) was focused on the sample by a microscope objective
($50\times$), having a spot diameter of about 1.5 $\mu$m. The PL
light was collected by the same objective, dispersed by a 1 m
monochromator and detected by a cooled CCD camera.

\begin{figure}
\includegraphics{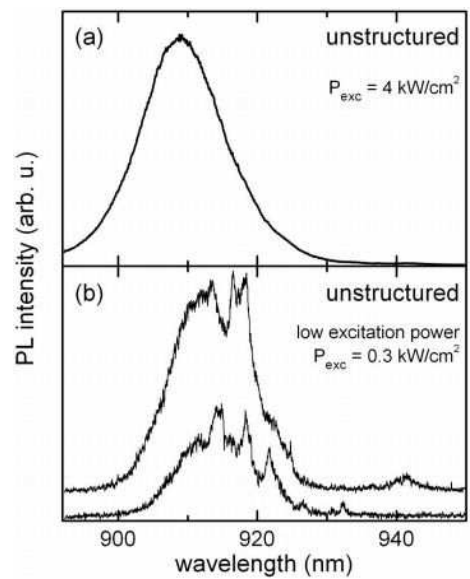}% Here is how to import EPS art
\caption{\label{fig2} PL spectra of the unstructured sample obtained
with different excitation power densities: (a) 4 kW/cm$^2$, (b) 0.3
kW/cm$^2$. The spectra in (b) (vertically shifted for clarity) were
obtained in one and the same measurement for slightly different
spatial positions within the area of the laser spot.}
\end{figure}

\begin{figure*}
\includegraphics[width=17.5cm]{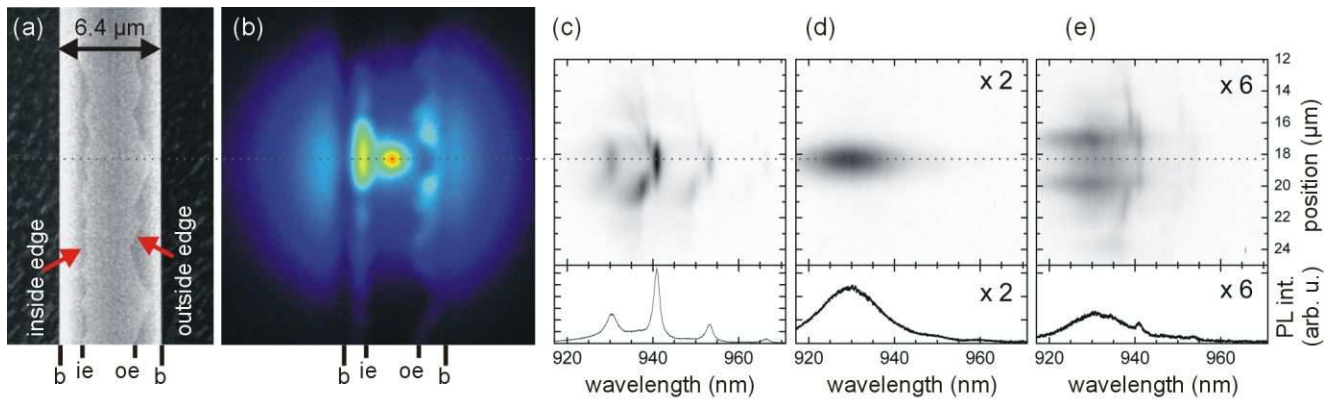}% Here is how to import EPS art
\caption{\label{fig3} (color online). (a) SEM picture of a part of
the microtube. Both, the inside and outside edge are visible. (b)
Undispersed PL image of the sample shown in (a). Both images are
equally scaled, the positions of the borders ($b$), the inside edge
($ie$) and the outside edge ($oe$) are marked. (c -- e) Spectrally
analyzed PL emision at the (c) inside edge, (d) radial position of
the laser spot, and (e) outside edge. In the upper panels, the
spatial information along the tube ($z$) axis is retained. The lower
panels show spectra obtained at the $z$ position on the level of the
laser spot (marked by the dotted horizontal line). The excitation
power density was about 5 kW/cm$^2$.}
\end{figure*}

Figure \ref{fig1} (a) shows a PL spectrum of the microtube. The
sequence of sharp peaks represent optical modes which are observed
for the first time in microtubes containing QWs. Before discussing
these modes in more detail we want to point out the excitation and
collection configuration used in this measurement. Here, the
positions of excitation and collection of the PL light were the same
in $z$ direction along the microtube axis, but could be shifted
against each other in radial direction of the microtube. This
situation is illustrated in the schematic inset in Fig. \ref{fig1}
by the two vertical arrows. The exciting laser generates electrons
and holes which quickly relax into the QW in the vicinity of the
focused laser spot on the microtube. There they form excitons which
then radiatively decay. A fraction of the emitted light does not
fulfill the condition of total reflection on the tube-wall/air
interface and therefore couples to leaky modes. The other fraction
does fulfill the condition of total reflection and is therefore
guided by the tube wall. Constructive interference leads to the
formation of optical resonator modes when, in a rather simple
picture, light is guided perpendicularly to the tube axis has the
same phase after one round trip. Therefore collecting the light at a
different position than exciting the QW suppresses the detection of
light out of leaky modes with respect to the cavity mode emission.
If we collect the PL light from the excitation position we can
identify the emission into leaky modes around 933 nm with a FWHM of
14 nm (23 meV). This is explicitly shown in Fig. \ref{fig3} (d)
which will be discussed later in the text. This signal is not
affected by interference inside the ring resonator and represents
the emission of a curved QW. It is red-shifted by about 20 nm (30
meV) compared to the QW emission of the unstructured sample, which
is shown in Fig. \ref{fig2} (a). The shifting is strain induced and
has been similarly reported by Hosoda \textit{et al.}
\cite{hosoda03}.

We now want to discuss the resonator modes in more detail. All modes
are linearly polarized with the electric field vector parallel to
the tube axis [TE polarization, Fig. \ref{fig1} (a)]. We find no
modes in perpendicular polarization, as can be seen in Fig.
\ref{fig1} (b). The strong peaks at about 939 nm, 951 nm, and 963.5
nm clearly show shoulders on their high energy side. Their origin is
not unambiguously clear. They might arise due to the lifting of the
degeneracy of a perfectly symmetric cylindrical resonator by the
inner and outer edges of the real structure (see sketch in Fig.
\ref{fig1}). Another explanation for the double peak structure might
be the existence of axially confined modes of slightly different
energy which spatially overlap or at least can not be separated by
our experimental setup. Such axially localized modes will be
thoroughly addressed later in this paper. At about 977 nm a further
mode with low intensity can be seen [marked by an arrow and
magnified by a factor of 20 in Fig. \ref{fig1} (a)]. By trying to
approximate the spectrum by a superposition of Lorentzians for the
resonant modes and their high-energy shoulders together with a
homogeneous Gaussian curve at about 933 nm with its above given FWHM
for the residual intensity from leaky modes, one can clearly
identify a further, rather broad mode on the high energy side of the
QW emission at about 929 nm. This peak can clearly be observed only
in the described excitation and detection configuration. Collecting
the PL light at the excitation position would strongly overlay this
peak by leaky modes.

Recapitulating, altogether, five linearly polarized optical modes
plus some shoulders are observed over an energy range of about 50 nm
(70 meV). These modes are reported here for the first time in
microtube ring resonators containing QWs as the optically active
material. The lowest lying mode is separated by about 44 nm (60 meV)
from the center QW emission energy. The observation of similar low
lying cavity modes have been reported for microdisks containing
ternary AlGaAs QWs and were attributed to the recombination of
electron-hole pairs localized in monolayer and alloy fluctuations
\cite{kipp04}. In fact, in measurements with low excitation power
density on our unstructured sample we observe single sharp lines
with spatially changing energy positions on the low energy side of
the QW emission, as expected for emission out of such localized
states. This can be seen in Fig. \ref{fig2} (b) where the excitation
power was decreased by more than a factor of 10 compared to the
measurements shown in (a) or in Fig. \ref{fig1}. Both spectra were
obtained in one and the same measurement for slightly different
spatial positions within the area of the laser spot. Regarding the
quality factors $Q=E/\Delta E$ one can deduce a weak trend of
slightly decreasing Q factors for increasing mode energies, still
below the center QW emission energy if one averages the values of
the main modes and its corresponding neighbors. This weak decreasing
can be explained with a slight increase of reabsorption of light
inside the cavity. Absorption is strongly enhanced for energies
above the QW emission center. Consequently the optical mode above
the QW emission energy experiences strong losses resulting in a
broadening by a factor of about ten compared to the sharpest peaks
on the low energy side.

Figure \ref{fig3} (a) shows a magnified SEM picture of a part of the
investigated self-supporting microtube bridge. As previously
mentioned, the tube has rolled-up slightly more than two times. For
this particular microtube the small region where the wall consists
of three rolled-up strained layers was orientated on top of the
tube, just like sketched in Fig. \ref{fig1}. Using a rather high
acceleration voltage of 20 kV at the SEM, not only the outside edge
of the microtube wall, but also the inside edge can be resolved.
Since the microtube wall in the region between the inside (left) and
outside (right) edge consists of three rolled-up layers it appears
slightly brighter in the SEM picture than the material besides this
region. We find that applying our preparation technique to the layer
system described above, both edges of the microtube tend to randomly
fray over some microns instead of forming straight lines. The
fraying occurs predominantly along the $\langle110\rangle$ direction
of the crystal, whereas the rolling direction of the microtube is
along $\langle100\rangle$. The spectrum in Fig.~\ref{fig1} proves
that optical modes still can develop despite the frayed edges since
it is actually obtained from exactly the microtube shown in Fig.
\ref{fig3} (a). By using the grating of the monochromator in zero
order as a mirror, we can directly image the microtube on the CCD
chip of the detector, as shown in Fig. \ref{fig3} (b). Here the
microtube was excited by the laser but in the collected signal the
laser stray light was cut off by an edge filter. Therefore Fig.
\ref{fig3} (b) shows an undispersed PL image of the
sample. We observe strong emission %only into leaky modes
at the excitation position centrally on the mircotube. We also
observe a large corona around this position due to PL emission of
the underlying GaAs substrate. The borders of the microtube become
apparent by vertical shadows. Furthermore both, the inside and
outside edge of the tube are visible in the CCD image. Even though
close to the resolution limit, the larger frays especially of the
outer edge which are clearly visible in the SEM picture in (a) can
also be identified in (b). Interestingly, we observe a strong
enhancement of PL emission near the inside edge of the microtube.
Having aligned the microtube axis in the way that its image is
parallel to the entrance slit of the spectrometer, we can spectrally
analyze the PL light of different radial positions retaining spatial
resolution along the tube axis. In Fig. \ref{fig3} (c) the signal
along the inside edge is analyzed. The vertical axis gives the
spatial position along the tube axis ($z$ direction), the horizontal
axis gives the spectral position, the PL intensity is encoded in a
grey scale. The lower panel in (c) shows a spectrum obtained at the
$z$ position on a level of the laser spot (dotted horizontal line).
The sequence of maxima of different wavelengths shows that the light
is indeed dominantly emitted out of resonant modes. In Fig.
\ref{fig3} (d) the signal emitted underneath the laser spot is
analyzed. In contrast to the inner edge, here, only one peak around
930 nm is observed, which is the emission of the QW into leaky modes
as already discussed in the context of Fig. \ref{fig1} (a). At last,
Fig. \ref{fig3} (e) analyzes the emission at the outside edge. Here,
we observe both, sequences of resonant modes and leaky modes, but
with a much smaller intensity than in (c) or (d). Note that the
intensity in (e) is multiplied by a factor of 6 compared to (c).
Thus, as a summary of Fig. \ref{fig3}, preferential emission of
modes at the inside edge is proved. This result, which is valid for
every position along the tube, indicates that the edges of the
microtube might be functionalized for a controlled or even
directional emission of light.

\begin{figure}
\includegraphics[width=8.5cm]{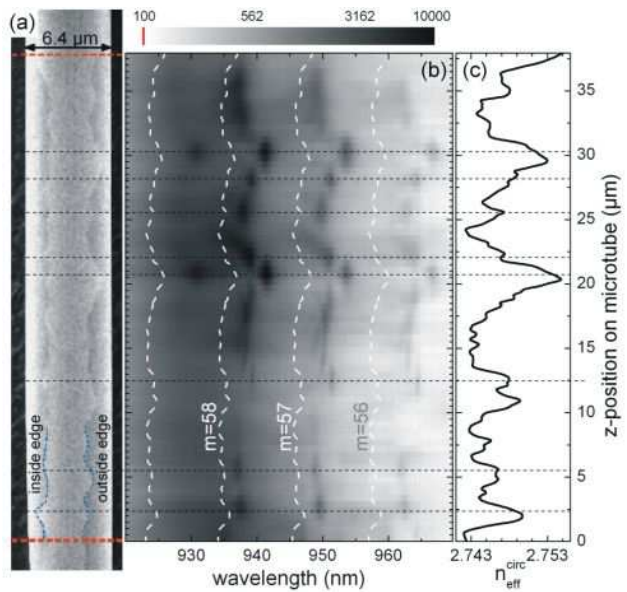}% Here is how to import EPS art
\caption{\label{fig4} (color online). (a) SEM picture of a part of
the microtube. Inside and outside edges are marked with broken lines
for clarification. The tube wall is thicker between these edges [see
sketch in Fig. \ref{fig1}]. (b) Scanning micro PL spectra show three
dimensionally confined optical modes. The white broken lines
represent calculated mode energies for $m =$ 59 to 56. (c)
Calculated effective refractive index vs. position. Horizontal
broken lines mark positions of some pronounced localized modes.}
\end{figure}

In order to investigate the influence of the frayed edges on the
mode spectrum, we performed micro PL measurements scanning along the
tube ($z$) axis which are shown in Fig. \ref{fig4} (b). The graph's
horizontal axis gives the wavelength of the detected light, whereas
its vertical axis gives the axial position on the microtube which
can directly be related to the SEM picture in Fig. \ref{fig4} (a).
The PL intensity is encoded in a gray scale where dark means high
signal. Note that the gray scale is chosen to be logarithmic in
order to better resolve in one and the same graph both, intense
peaks on a comparatively intense background and small peaks on a
weak background. The microtube was scanned in 80 steps over a length
of 35 $\mu$µm [between the two broken horizontal lines in Fig.
\ref{fig4} (a)]. In $z$ direction the collecting position of the PL
light was centered on the excitation spot, in radial direction we
collected over the whole microtube.

For nearly every position on the microtube we observe three or four
optical modes. Their energies are shifting along the $z$ direction,
but interestingly this shifting is not continuous: The mode energies
sometimes seem to be spatially pinned. First, we want to explain the
shifting by an expanded model of a closed waveguide. For each
position in $z$ direction we regard a cross-section perpendicular
the tube axis as a circular {waveguide}, having a diameter of
$d=6.4~\mu$m. The waveguide thickness abruptly changes at the edges
from 130 nm to 195 nm. For each region we translate the thickness in
a related effective refractive index $n_{\mathrm{eff}}^{130nm}$ and
$n_{\mathrm{eff}}^{195nm}$ respectively \cite{kipp06}. For this
calculation we assumed the average refractive index of the layer
system to be $n(E)=0.3225\times E[\mathrm{eV}] + 2.98232$ by
linearly approximating values given in Refs.~\onlinecite{takagi78}
and \onlinecite{Pikhtin80}. From Fig. \ref{fig4} (a) we determine
the distance $L^{195nm}(z)$ from the inside to the outside edge on
the tube surface by taking into account that the picture is a
projection of a curved surface. We then can define an overall
effective refractive index for the whole circular waveguide for each
$z$ position: $n_{\mathrm{eff}}^{\mathrm{circ}}(z)=
n_{\mathrm{eff}}^{195nm} L^{195nm}(z)/(\pi d) +
n_{\mathrm{eff}}^{130nm}[1 - L^{\mathrm{195 nm}}(z)/(\pi d)]$. The
periodic boundary condition for a resonant mode then reads
$n_{\mathrm{eff}}^{\mathrm{circ}}(z) \pi d = m \lambda$, with the
vacuum wavelength $\lambda$ and the azimuthal mode number $m \in
\mathds{N}$. The resonances calculated with this model have
azimuthal mode numbers around $m=57$ and are depicted in Fig.
\ref{fig4} (b) as bright broken lines. The absolute position of the
calculated $m$th mode strongly depends on the exact dimension of the
microtube, whereas the calculated mode spacing fits quite well to
the measurements. Deviations can be explained by slightly
insufficient approximated refractive index of the layer system,
especially because of neglecting absorption inside the QW. As a
striking result, the calculation reveals that the overall $z$
dependency of the measured resonant wavelengths is nicely
approximated by our model.

In the following we want to concentrate on the spatially pinned
resonances. These resonances can be seen in Fig. \ref{fig4} (b) as
isolated dark spots with no significant wavelength shift over one or
more microns. This implies that these resonances are localized
modes, confined also in $z$ direction along the tube axis. Comparing
Fig. \ref{fig4} (a) and (b) it seems that some of these localized
modes can be attributed to positions on the tube with a broad
distance between inside and outside edge. To work out this point in
more detail, Fig. \ref{fig4} (c) shows
$n_{\mathrm{eff}}^{\mathrm{circ}}(z)$ calculated for $\lambda = $
940 nm, which essentially reflects the distance between the edges in
dependency of the $z$ position. Comparing now (b) and (c), it
becomes obvious that optical modes are predominantly localized in
regions of local maxima of $n_{\mathrm{eff}}^{\mathrm{circ}}(z)$
representing local maxima of $L^{195nm}(z)$. To better visualize
this behavior, all pronounced localized modes are indicated with
horizontal lines in Fig. \ref{fig4} (b). Confinement of light along
the $z$ direction is achieved by a change of
$n_{\mathrm{eff}}^{\mathrm{circ}}(z)$. This important result can be
qualitatively explained within the waveguide model. Until now we
implied in our model light having no wave vector component along the
tube axis, because, for a perfect infinite microtube, this light
would just run away along the tube axis. If we now regard light with
a finite but small wave vector component along the axis, this light
can experience total internal reflection also in $z$ direction,
which leads to a three dimensional confinement. Total internal
reflection in this case is quite similar to the situation in a
graded-index optical fiber. Our model using an averaged effective
refractive index for a circular cross-sectional area of a microtube
explains our measured data astonishingly well, despite this model
neglects the abrupt changes of the inside and outside diameter. The
experimental data shows that it is possible to confine light also
along the axis of a microtube ring resonator by a slight change of
the wall geometry along the axis. This result opens up new vistas
for a controlled three dimensional confinement of light inside a
microtube ring resonator by the use of a lithographically controlled
variation of the inside and outside edge of the strained layer
system before the roll-up process takes place. In this context we
would like to refer to a recent theoretical work by Louyer \emph{et
al.} on prolate-shaped dielectric microresonators \cite{louyer05}.
We believe that our technique of laterally structuring the edges of
a microtube before the roll-up process, which is presented in this
paper, might make it possible to fabricate prolate-shaped microtube
resonators with similar characteristics as described in Ref.
\onlinecite{louyer05}.

In summary, we report on a microtube resonator with embedded QWs. We
observe optical modes in a quite large energy range which are partly
affected by reabsorption. Scanning micro PL measurements demonstrate
(i) preferential emission of light near the inside edge of the
microtube and (ii) confinement of light also in direction along the
tube axis. The confinement is induced by spatial variations of the
inside and outside edges of a microtube along its axis and can be
approximated in an expanded waveguide model. The presented results
highlight the possibility of both, a controlled emission and a
controlled three dimensional confinement of light in a microtube by
a lithographically controlled definition of the microtube edges.

We gratefully acknowledge financial support of the Deutsche
Forschungsgemeinschaft via the SFB 508 ``Quantum Materials'' and the
Graduiertenkolleg 1286 ``Functional Metal-Semiconductor Hybrid
Systems''.

%\newpage %Just because of unusual number of tables stacked at end
%\bibliography{../../../../bibtexfiles/tobbib}% Produces the bibliography via BibTeX.
%\bibliography{tobbibkurz}% Produces the bibliography via BibTeX.
\end{document}